\documentclass[conference]{IEEEtran}
\IEEEoverridecommandlockouts

\usepackage{cite}
\usepackage{amsmath,amssymb,amsfonts}
\usepackage{bm}
\usepackage{algorithmic}
\usepackage{graphicx}
\usepackage{textcomp}
\usepackage{xcolor}
\newcommand{\argmax}{\mathop{\rm arg~max}\limits}
\newcommand{\argmin}{\mathop{\rm arg~min}\limits}
\def\BibTeX{{\rm B\kern-.05em{\sc i\kern-.025em b}\kern-.08em
    T\kern-.1667em\lower.7ex\hbox{E}\kern-.125emX}}
\begin{document}

\title{Few-shot Personalization via In-Context Learning for Speech Emotion Recognition \\based on Speech-Language Model}

\author{
\IEEEauthorblockN{Mana Ihori, Taiga Yamane, Naotaka Kawata, Naoki Makishima,\\ Tomohiro Tanaka, Satoshi Suzuki, Shota Orihashi, Ryo Masumura}
\IEEEauthorblockA{\textit{Human Informatics Laboratories, NTT, Inc.}, Japan}
}


\maketitle

\begin{abstract}
This paper proposes a personalization method for speech emotion recognition (SER) through in-context learning (ICL). Since the expression of emotions varies from person to person, speaker-specific adaptation is crucial for improving the SER performance. Conventional SER methods have been personalized using emotional utterances of a target speaker, but it is often difficult to prepare utterances corresponding to all emotion labels in advance. Our idea to overcome this difficulty is to obtain speaker characteristics by conditioning a few emotional utterances of the target speaker in ICL-based inference. ICL is a method to perform unseen tasks by conditioning a few input-output examples through inference in large language models (LLMs). We meta-train a speech-language model extended from the LLM to learn how to perform personalized SER via ICL. Experimental results using our newly collected SER dataset demonstrate that the proposed method outperforms conventional methods.
\end{abstract}

\begin{IEEEkeywords}
speech emotion recognition, in-context learning, speech-language model, meta-training for in-context learning
\end{IEEEkeywords}

\section{Introduction}
Speech emotion recognition (SER) plays an important role in effectively understanding human-computer interaction and human communication.
For example, SER is expected to be applied to applications such as estimating customer satisfaction in contact centers \cite{8960433} and developing speech dialogue systems that can empathize and sympathize \cite{10.5555/1620932.1620941}.
On the other hand, the expression of emotions varies depending on various individual characteristics such as culture, environment, and personality, making SER a difficult problem to generalize \cite{larsen1987affect,russell1994there,gross1997emotion,chaplin2013gender,sherman2015independent}.
In particular, the SER performance significantly degrades when a speaker who is not included in the training data is input during inference \cite{9836236,doddington1998sheep}.
Therefore, SER is a difficult task to adapt to unseen speakers.

Conventionally, to improve the SER performance for unseen speakers, methods using enrollment utterances, which are pre-collected and reflect speaker-specific emotional traits, have been proposed \cite{9767771,9428217,triantafyllopoulos2022,triantafyllopoulos2024}.
In these methods, enrollment utterances (e.g., neutral or each emotional utterance) are input into the SER model simultaneously with the target speech, and the utterances are used as guidance for speaker adaptation.
In addition, it is reported that the method using enrollment utterances corresponding to all emotion labels has outperformed the method using only a neutral utterance \cite{triantafyllopoulos2024}.
In these personalization methods, the SER models have been trained using a fixed number of enrollment utterances with fixed emotion labels.
Thus, it is necessary to prepare enrollment utterances with the same emotion labels as those used during training for unseen speakers at inference time.
However, in practice, it is difficult to always prepare enrollment utterances with the same emotion labels as those used during training because the enrollment utterances of the target speaker may have imbalanced emotion labels or may not be available.

 \begin{figure}[t]
  \centering
  \centerline{\includegraphics[clip, width=9cm]{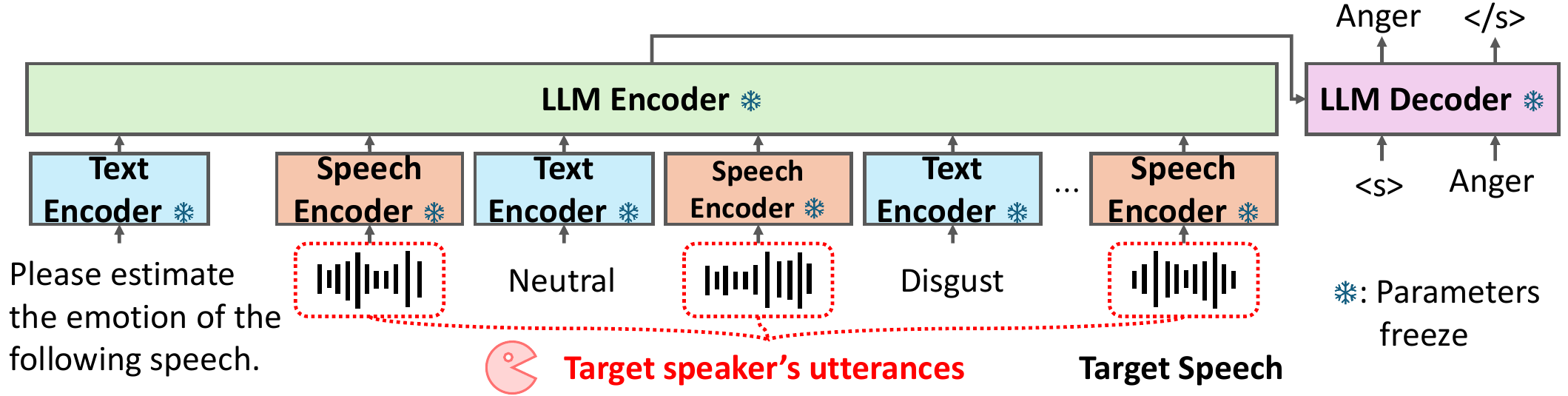}}
  \caption{Example of personalization for SER via ICL in speech LM.}
  \label{fig:icl}
\end{figure}

To perform personalization for SER using an arbitrary number of enrollment utterances with any emotion labels, we focus on in-context learning (ICL) in large language models (LLMs).
ICL is a method for adapting LLMs to unseen tasks by conditioning a few input-output examples through inference without updating their parameters \cite{NEURIPS2020_1457c0d6}.
In ICL, we can flexibly adjust the number and types of examples and perform zero-shot inference without using examples.
In this study, we consider performing ICL not to adapt the SER model to unseen tasks, but to adapt it to unseen speakers.
By conditioning the SER model on a few enrollment utterance-label pairs uttered by a target speaker, we expect the model to acquire the speaker characteristics necessary for speaker-adaptive SER based on the correspondence between these utterances and these labels.
In addition, the model can improve SER performance for unseen speakers by personalizing SER through ICL-based inference.
Since the number of enrollment utterances and their emotion labels uttered by a target speaker do not need to be fixed in ICL, dynamic SER personalization becomes possible.

In this paper, we propose a method to perform personalized SER for unseen speakers by leveraging the ICL capability of LLMs.
In the proposed method, we construct a speech-language model (speech LM) that can perform dynamic speaker adaptation for SER through ICL-based inference by conditioning a few enrollment utterance-label pairs uttered by a target speaker, as shown in Fig. \ref{fig:icl}.
The speech LM is constructed by extending the LLM to enable cross-modal input of speech and text, and conventional speech LMs for SER have not taken into account speaker characteristics, which is important for SER \cite{zhao2025steeringlanguagemodelstable,xu2024secap}.
To acquire speaker characteristics of unseen speakers using the ICL capability of the LLM, we meta-train the speech LM using meta-training for ICL (MetaICL), which is a fine-tuning method for training how to perform ICL \cite{min-etal-2022-metaicl}.
Specifically, we prepare multiple enrollment utterances of various speakers and fine-tune the LLM to predict other emotions given partial emotions of the target speaker in MetaICL.
By modeling in this way, we expect that the speech LM can perform robust personalized SER for unseen speakers via ICL.
Experimental results using a newly collected SER dataset demonstrate that the proposed method improves unweighted accuracy (\%) by 2–-3 pt regardless of which enrollment utterances were used up to the 2-shot setting and up to 8.2 pt compared to performing inference without ICL in the speech LM.

\section{Related work}
\paragraph*{SER}
With the advancement of deep learning, neural networks are generally used to predict emotion labels given speech features in the SER task \cite{10089511}.
In SER, there are two types of settings: speaker-dependent settings, which use the same speaker for training and inference \cite{8639583,9054629,jalal2020removing}, and speaker-independent settings, which use different speakers for training and inference \cite{schuller2005speaker,1521560}.
In speaker-dependent settings, it has been reported that recognition performance can be improved by constructing a speaker-dependent SER model, and speaker information is considered to be an important factor in SER \cite{6854514}.
In the speaker-independent setting, by using speaker information such as age \cite{7953138}, gender \cite{zhang2018gender,li2019improved}, and speaker vectors \cite{le2021speaker,10753506}, SER performance has been improved.
These methods demonstrated that speaker information is effective for SER, but it remains unclear which characteristics of speaker information are particularly effective.
In addition, methods for removing speaker individuality have been proposed to perform generalized SER \cite{9747460}.
However, this method embeds individual differences in speakers into a common space, which may reduce the accuracy of SER for speakers with large individual differences.
Furthermore, SER methods using the enrollment utterances, which are pre-collected and reflect speaker-specific emotional traits, have been proposed \cite{9767771,9428217,triantafyllopoulos2022,triantafyllopoulos2024}.
In these methods, the model combines the emotional information in enrollment utterances with the target speech to improve SER performance.
This paper aims to perform personalized SER for unseen speakers by using enrollment utterances and their labels in ICL.

\paragraph*{Speech LM}
Recently, many speech LMs that enable cross-modal input speech and text have appeared by extending the LLM, and these models can perform various speech tasks, including SER \cite{tang2023salmonn,kong2024audio,Qwen2-Audio}.
Also, speech LMs specialized for the SER task have appeared.
In these models, \cite{zhao2025steeringlanguagemodelstable} proposes a method to suppress hallucinations of emotion labels, and \cite{xu2024secap} proposes a method to generate emotion captions rather than to output emotion labels corresponding to the input speech.
However, these methods do not leverage speaker characteristics such as age, gender, and enrollment utterances.
With the success of ICL in LLMs, ICL is also performed in speech LMs.
In ICL for speech LMs, it has been reported that performance can be improved by performing ICL-based inference for tasks such as automatic speech recognition (ASR) and speech translation \cite{chen2024salm,10446502,roll2025incontext}.
In addition, there are several studies that perform unseen tasks using ICL-based inference in the speech LM.
For example, \cite{pan2024cosmic} showed that speech LMs can perform speech translation of combinations that have not been learned during inference using ICL by learning mainly speech recognition and speech translation tasks.
Also, \cite{chang2024} showed that unseen tasks can be performed through ICL by training various speech understanding tasks.
In this study, we focus on SER, which is more speaker-dependent than ASR, and investigate whether it is possible to apply the speech LM to unseen speakers rather than to unseen tasks by using a few enrollment utterance-label pairs of a target speaker in ICL.

 \begin{figure*}[t]
  \centering
  \centerline{\includegraphics[clip, width=18cm]{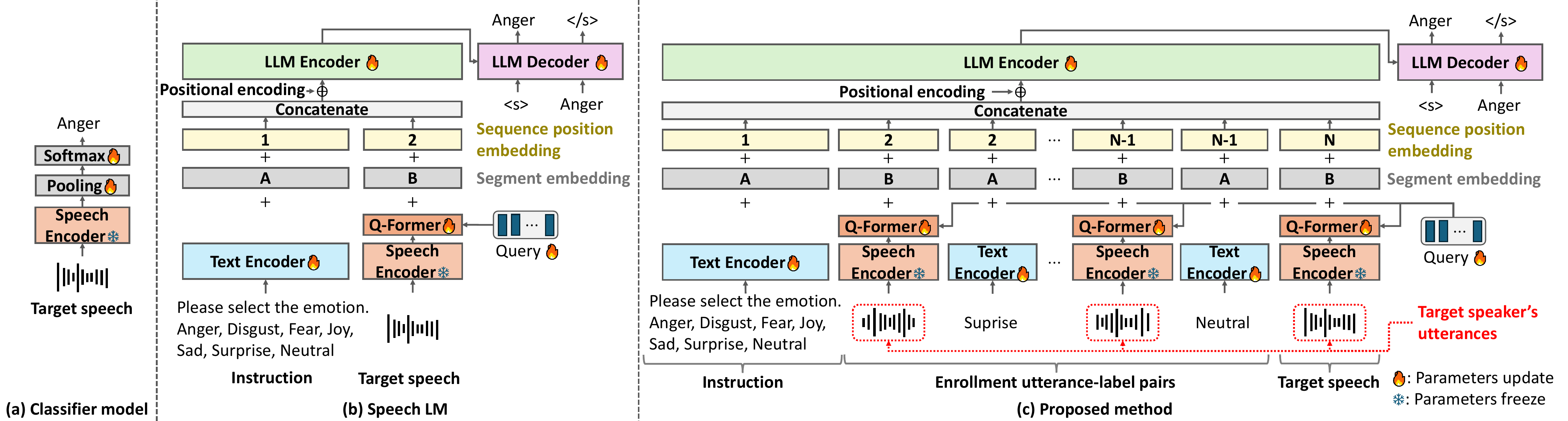}}
  \caption{Overview of classifier model, speech language model, and proposed method for SER.}
  \label{fig:main}
\end{figure*}

\section{Preliminaries}
In this section, we define the modeling method of a classifier model and speech LM for SER used in this paper. 

\subsection{Classifier model}
In this paper, the classifier model predicts an emotion label $y \in \{y_1,\cdots, y_N\}$, where $N$ is the number of emotion labels, given speech features $\bm{X} = \{x_1, \cdots, x_m, \cdots, x_M\}$, where $x_m$ is the $m$-th speech feature and $M$ is the number of speech features, as shown in Fig. \ref{fig:main} (a).
The probability of $y$ is defined as $P(y|\bm{X};\bm{\Theta}_{\rm cls})$, where $\bm{\Theta}_{\rm cls}$ is a trainable parameter set.
$P(y|\bm{X};\bm{\Theta}_{\rm cls})$ is computed with a transformer-based speech encoder and an output layer.

\paragraph*{Network Structure}
The speech encoder converts $\bm{X}$ into speech representations $\bm{H}$ as
\begin{equation}
    \bm{H} = {\rm TransformerEnc}(\bm{X}; \theta_{\rm speech}),
\end{equation}
where $\rm{TransformerEnc}()$ is a function of the transformer encoder blocks that consist of multi-head self-attention layers and position-wise feed-forward networks \cite{vaswani2017attention}, and $\theta_{\rm speech} \in \bm{\Theta}_{\rm cls}$ are the trainable parameters.
Next, the attentive pooling converts variable-length hidden vectors $\bm{H}$ into a fixed-size vector $\bm{h}$ as
\begin{equation}
    \bm{h} = {\rm AttentivePooling}(\bm{H};\theta_{\rm{pool}}),
\end{equation}
where $\rm{AttentivePooling}()$ is the attentive pooling function, and $\theta_{\rm{pool}} \in \bm{\Theta_{\rm cls}}$ are the trainable parameters. 
The probability of $y$ is calculated by
\begin{equation}
    P(y|\bm{X};\bm{\Theta}_{\rm cls}) = {\rm Softmax}(\bm{h};\theta_{\rm{soft}}),
\end{equation}
where $\rm{Softmax}()$ is a softmax-activation layer with a linear transformation, and $\theta_{\rm soft} \in \bm{\Theta}_{\rm cls}$ are the trainable parameters.

\paragraph*{Training}
The model parameter set $\bm{\Theta}_{\rm cls}$ is optimized from the training dataset ${\cal D}_{\rm cls}$ $=$ $\{(\bm{X}^1,$ $y^1),$ $\cdots,$ $(\bm{X}^{|{\cal D}_{\rm cls}|},$ $y^{|{\cal D}_{\rm cls}|})\}$ as
\begin{equation}
    \hat{\bm{\Theta}}_{\rm cls} = \argmin_{\bm{\Theta_{\rm cls}}} - \sum_{(\bm{X}, y) \in {\cal D}_{\rm cls}} \log P(y|\bm{X}; \bm{\Theta}_{\rm cls}).
\end{equation}

\paragraph*{Inference}
The inference of $y$ is defined as
\begin{equation}
    \hat{y} = \argmax_{y} P(y|\bm{X};\hat{\bm{\Theta}}_{\rm cls}).
\end{equation}

\subsection{Speech LM}
In this paper, the speech LM predicts an emotion text $\bm{Z} = \{z_1, \cdots, z_t, \cdots, z_T\}$, where $z_t$ is the $t$-th token and $T$ is the number of tokens in the emotion text, given instruction text $\bm{W}$ and $\bm{X}$, as shown in Fig. \ref{fig:main} (b).
The generation probability of $\bm{Z}$ is defined as
\begin{equation}
    P(\bm{Z}|\bm{W},\bm{X};\bm{\Theta}_{\rm sl}) = \prod_{t=1}^T P(z_t|z_{1:t-1}, \bm{W}, \bm{X}; \bm{\Theta}_{\rm sl}),
\end{equation}
where $z_{1:t-1} = \{z_1, \cdots, z_{t-1}\}$ and $\bm{\Theta}_{\rm sl}$ is a trainable parameter set.
$P(\bm{Z}|\bm{W},\bm{X};\bm{\Theta}_{\rm sl})$ is computed with the speech LM that combines a speech encoder with an LLM.
In this paper, we use an instruction-tuned encoder-decoder type LLM, and the speech LM consists of a speech encoder, a conversion module, a text encoder, an LLM encoder, and an LLM decoder.

\paragraph*{Instruction-tuned LLM}
In this paper, we construct a speech LM by extending an instruction-tuned LLM.
Instruction-tuning is a fine-tuning method to improve the ability to respond according to instructions and the ICL capability \cite{weifinetuned}, and the LLM is fine-tuned to predict a response text $\bm{O}$ given an instruction text $\bm{W}$.
The probabilities of $\bm{O}$ are defined as $P(\bm{O}|\bm{W};\bm{\Theta}_{\rm llm})$, where $\bm{\Theta}_{\rm llm}$ is a trainable parameter set, which is initialized by pretraining with a next token prediction objective. 
$\bm{\Theta}_{\rm llm}$ is optimized from the training dataset ${\cal D}_{\rm ins} = \{(\bm{W}^1, \bm{O}^1), \cdots, (\bm{W}^{|{\cal D}_{\rm ins}|},\bm{O}^{|{\cal D}_{\rm ins}|})\}$ as
\begin{equation}
    \hat{\bm{\Theta}}_{\rm llm} = \argmin_{\bm{\Theta}_{\rm llm}} \sum_{(\bm{W}, \bm{O}) \in {\cal D}_{\rm ins}} P(\bm{O}|\bm{W};\bm{\Theta}_{\rm llm}).
\end{equation}

\paragraph*{Speech encoder}
In the speech encoder, $\bm{X}$ is converted into $\bm{H}$ as shown in Eq. (1), and $\theta_{\rm speech}$ is fixed.

\paragraph*{Conversion module}
To input $\bm{X}$ into the LLM, a conversion module is required to align the input embedding space of the LLM and the speech representation.
In this paper, we use Q-Former \cite{li2023blip}, which can convert input sequences of arbitrary length into fixed-length representations.
In the Q-Former, fixed-length query representation $\bm{U}$ is outputted given the query that is trainable fixed-length embedding $\bm{Q}$ and $\bm{H}$ as
\begin{equation}
    \bm{U} = {\rm QFormer}(\bm{Q}, {\bm{H}; \theta_{\rm qf}}),
\end{equation}
where ${\rm QFormer}()$ is a function of the transformer decoder blocks without a causal attention mask that consist of multi-head self-attention and cross-attention layers and position-wise feed-forward networks \cite{vaswani2017attention}, and $\theta_{\rm qf} \in \bm{\Theta}_{\rm sl}$ are the trainable parameters.

\paragraph*{Text encoder}
In the text encoder, $\bm{W}$ is converted into a continuous representation $\bm{R}$ as
\begin{equation}
    \bm{R} = {\rm Embedding}(\bm{W};\hat{\bm{\Theta}}_{\rm llm}),
\end{equation}
where ${\rm Embedding}()$ is the function that converts a token into a continuous vector and $\hat{\bm{\Theta}}_{\rm llm} \in \bm{\Theta}_{\rm sl}$.

\paragraph*{LLM encoder}
In the LLM encoder, $\bm{R}$ and $\bm{U}$ are converted hidden representation $\bm{C}$ as
\begin{equation}
    \bm{C} = {\rm TransformerEnc}(\bm{A}; \hat{\bm{\Theta}}_{\rm llm}),
\end{equation}
\begin{equation}
    \bm{A} = {\rm PositionalEncoding}([\bar{\bm{R}}^T, \bar{\bm{U}}^T]),
\end{equation}
\begin{equation}
    \bar{\bm{R}} = {\rm SeqPosEmb}({\rm SegmentEmb}(\bm{R}; \theta_{\rm seg}); \theta_{\rm seq}),
\end{equation}
\begin{equation}
    \bar{\bm{U}} = {\rm SeqPosEmb}({\rm SegmentEmb}(\bm{U}; \theta_{\rm seg}); \theta_{\rm seq}),
\end{equation}
where ${\rm PosEnc}()$ is positional encoding, ${\rm SeqPosEmb}()$ is sequence position embedding that indicates the order of the input sequences, ${\rm SegmentEmb}()$ is segment embedding that distinguishes different modalities and $\{\theta_{\rm seg}, \theta_{\rm seq}\} \in \bm{\Theta}_{\rm sl}$ are the trainable parameters.

\paragraph*{LLM decoder}
In the LLM decoder, the generation probability of $z_t$ is calculated given $\bm{C}$ and $z_{1:t-1} = \{z_1, \cdots, z_{t-1}\}$ as
\begin{equation}
    P(z_t|z_{1:t-1},\bm{W},\bm{X};\bm{\Theta}_{\rm sl}) = {\rm Softmax}(\bm{F}; \hat{\bm{\Theta}}_{\rm llm}),
\end{equation}
\begin{equation}
    \bm{F} = {\rm TransformerDec}(\bm{C}, z_{1:t-1};\hat{\bm{\Theta}}_{\rm llm}),
\end{equation}
where ${\rm TransformerDec}()$ is a function of the transformer decoder blocks.

\paragraph*{Training}
The model parameter set $\bm{\Theta}_{\rm sl}$ is optimized from training dataset ${\cal D_{\rm text}} = \{(\bm{X}^1, \bm{Z}^1), \cdots, (\bm{X}^{|{\cal D}_{\rm text}|}, \bm{Z}^{|{\cal D}_{\rm text}|})\}$ as
\begin{equation}
    \hat{\bm{\Theta}}_{\rm sl} = \argmin_{\Theta_{\rm sl}} - \sum_{(\bm{X},\bm{Z}) \in {\cal D}_{\rm text}} P(\bm{Z}|\bm{W},\bm{X};\bm{\Theta}_{\rm sl}).
\end{equation}
$\bm{W}$ is a common instruction text across ${\cal D}_{\rm text}$.
During training, the speech encoder parameters are fixed, and the LLM parameters are updated so that the speech representation approaches the LLM embedding space.

\paragraph*{Inference}
The inference of $\bm{Z}$ is defined as
\begin{equation}
    \hat{\bm{Z}} = \argmax_{\bm{Z}} P(\bm{Z}|\bm{W},\bm{X};\hat{\bm{\Theta}}_{\rm sl}).
\end{equation}

\section{Proposed Method}
\subsection{Strategy of personalization via ICL}
This paper proposes a method to perform personalized SER for unseen speakers via ICL.
Fig. \ref{fig:main} (c) shows the overview of the proposed method.
We expect the speech LM extended from an LLM with ICL capability to perform personalization for a speaker through inference by conditioning a few enrollment utterance-label pairs uttered by the target speaker in ICL.
In the proposed method, we construct the speech LM that can be adapted to unseen speakers through ICL by fine-tuning the LLM that can be adapted to unseen tasks through ICL.
In MetaICL, the speech LM is fine-tuned in two stages.
In the first stage, since training the model using long sequences including enrollment utterances may cause unstable learning, the model is fine-tuned without the enrollment utterance-label pair, as in Eq. (15).
In the second stage, the model is meta-trained to generate emotion text given an instruction text, a few enrollment utterance-label pairs, and target speech features.
At that stage, zero to seven enrollment utterance-label pairs are randomly selected from the dataset of the same speaker as the target speech, satisfying the following settings ($\text{TU+LU}_{0:7}$).
\begin{itemize}
    \item \textbf{Target-Uncontrolled (TU)}: The same emotion as the target appears or not in the enrollment utterance-label pairs.
    \item \textbf{Label-Uncontrolled (LU)}: The enrollment utterance-label pairs have the same or different emotions.
\end{itemize}
By using enrollment utterance-label pairs selected under these settings in MetaICL, we expect the speech LM to perform ICL even when utterances with any emotion labels are used.

In addition, to investigate whether performance changes depending on the combination of emotions, we select enrollment utterance-label pairs from the above and the following settings in ICL-based inference.
\begin{itemize}
    \item \textbf{Target-Overlap (TO)}: The same emotion label as the target appears in the enrollment utterance-label pairs.
    \item \textbf{Target-Exclusive (TE)}: All enrollment utterance-label pairs are different from the target emotion.
    \item \textbf{Label-Overlap (LO)}: All enrollment utterance-label pairs have the same emotion emotion.
    \item \textbf{Label-Disjoint (LD)}: All enrollment utterance-label pairs have different emotion emotions.
\end{itemize}

\subsection{MetaICL}
In MetaICL, the speech LM in Eq. (6) is fine-tuned using $k$-shot enrollment utterance-label pairs $\bm{S} = \{(\bm{X}^1, \bm{Z}^1), \cdots, (\bm{X}^k, \bm{Z}^k)\}$ by optimizing a trainable parameter set $\bm{\Theta}_{\rm meta}$, which is initialized from $\hat{\bm{\Theta}}_{\rm sl}$ in Eq. (15).
$\bm{\Theta}_{\rm meta}$ is optimized using a set of datasets ${\cal D}=\{{\cal D}_1, \cdots, {\cal D}_j, \cdots, {\cal D}_J\}$, where $J$ is the number of speakers.
The $t$-th dataset ${\cal D}_j = \{(\bm{X}_j^1, \bm{Z}_j^1), \cdots, (\bm{X}_j^{|{\cal D}_j|},\bm{Z}_j^{|{\cal D}_j|})\}$ has a set of enrollment utterance-label pair uttered by the $j$-th speaker.
Thus, $\bm{\Theta}_{\rm meta}$ is optimized using ${\cal D}$ as 
\begin{align}
    & \hat{\bm{\Theta}}_{\rm meta} = \\
    & \argmin_{\bm{\Theta}_{\rm meta}} - \sum_{j=1}^J \sum_{(\bm{X}_j,\bm{Z}_j) \in {\cal D}_j} \log P(\bm{Z}_j|\bm{W}, \bm{S}_j, \bm{X}_j;\bm{\Theta}_{\rm meta}), \notag
\end{align}
\begin{equation}
    \bm{S}_j = {\rm SelectProcedure}(\bm{X}_j,\bm{Z}_j, k, {\cal D}_j),
\end{equation}
where ${\rm SelectProcedure}()$ randomly selects $k$ samples from the set excluding $\bm{X}_j$ and $\bm{Z}_j$ from ${\cal D}_j$ in accordance with TU+LU setting.

\subsection{Personalization via ICL}
In ICL-based inference, the speech LM personalizes SER by conditioning $k$-shot enrollment utterance-label pairs uttered by a target speaker.
In $k$-shot ICL-based inference, the inference of $\bm{Z}$ is defined as
\begin{equation}
    \hat{\bm{Z}} = \argmax_{\bm{Z}} P(\bm{Z}|\bm{W}, \bm{S}_{\rm tgt}, \bm{X}_{\rm tgt};\hat{\bm{\Theta}}_{\rm meta}),
\end{equation}
where $\bm{S}_{\rm tgt}=\{(\bm{X}^1_{\rm tgt}, \bm{Z}^1), \cdots, (\bm{X}^k_{\rm tgt}, \bm{Z}^k)\}$, and $\bm{X}_{\rm tgt}^k$ is the $k$-th speech features uttered by a target speaker.
In addition, the speech LM can predict $\bm{Z}$ without giving $\bm{S}_{\rm tgt}$, which is called ``zero-shot inference'', when $\hat{\bm{\Theta}}_{\rm meta}$ is used in Eq. (17).

\begin{table}[tb]
  \centering
  \caption{Number of utterances in training, validation, and test sets in our created dataset. Each speaker has 50 utterances for seven emotions.}
  \begin{tabular}{lrrrr}
      \hline
       Gender & Age group & \multicolumn{3}{c}{Number of utterances} \\
        & & Train & Valid & Test \\
      \hline
      Male & 10s &  1,050 & 700 & 700 \\
       & 20s & 24,850 & 3,150 & 3,150  \\
       & 30s & 14,000 & 1,400 & 1,400  \\
       & 40s & 23,100 & 2,800 & 2,800 \\
       & 50s & 22,750 & 2,800 & 2,800 \\
       & 60s & 17,150 & 2,100 & 2,100 \\
     \hline
     Female & 10s & 1,050 & 700 & 700\\
      & 20s & 24,850 & 2,800 & 2,800 \\
      & 30s & 17,850 & 2,100 & 2,100\\
      & 40s & 28,000 & 3,150 & 3,150 \\
      & 50s & 28,000 & 3,150 & 3,150 \\
      & 60s & 22,050 & 2,450 & 2,450 \\
      & 70s & 350 & 0 & 350\\
      \hline
      Total & & 225,050 & 27,300  & 27,650 \\
      \hline
  \end{tabular}
  \label{table:data}
\end{table}

\section{Experiments}
\subsection{Dataset}
To evaluate personalized SER for unseen speakers, multiple emotional utterances of various speakers are required.
However, the dataset used in conventional SER only includes about 10 speakers \cite{busso2008iemocap,burkhardt2005database,haq2009speaker}, so it is difficult to evaluate in the speaker-independent setting with a sufficient number of speakers.
Thus, this study newly collected a Japanese SER dataset that has many speakers.
This dataset consists of 800 native Japanese speakers (368 males and 432 females) who performed 50 utterances each for seven emotions: anger, disgust, fear, joy, sadness, surprise, and neutral.
The text read aloud is unrelated to the speaker's emotional state, so even if the text content is taken into consideration, it does not contribute to emotion estimation.
We instructed the speakers to ``speak as clearly as possible so that your emotions come across'' for each text, and recorded speech using a headset microphone and a laptop computer in a conference room free of background noise.
We divided this dataset into training, development, and test sets so that the gender ratio and age groups are evenly distributed, as shown in Table \ref{table:data}.
In this dataset, we used a common instruction text: ``Please select the appropriate emotion for the input speech from the following: Neutral, Surprise, Sadness, Joy, Fear, Disgust, Anger.''

\subsection{Models}
We compare the performance of four classifier models, two speech LMs, and the proposed method.
In the classifier model and speech LM, we utilized an in-house transformer-based speech encoder, which has 42M parameters, and these parameters were pre-trained in various speech-understanding tasks.
In the speech LM, we utilized an encoder-decoder style in-house LLM, which has 0.6B parameters.
The parameters were pre-trained with a large amount of text based on unifying language learning paradigms \cite{tayul2}, and instruction-tuned with various language tasks.
Details of the comparison method are shown below.
\paragraph*{Classifier model}
\begin{itemize}
    \item \textbf{Scratch model}: a transformer-based model in which all parameters were randomly initialized, defined as Eq. (4),
    \item \textbf{Pretrained model}: a transformer-based model with the pretrained speech encoder.  In this model, the parameters of the speech encoder were fixed.
    \item $\textbf{Personalized model}_N$ \cite{9428217}: a model with a single neutral utterance added to the input of the pretrained model.
    \item $\textbf{Personalized model}_A$ \cite{triantafyllopoulos2024}: a model with each emotional utterance of a target speaker added to the input of the pretrained model.
\end{itemize}

\paragraph*{Speech LM}
\begin{itemize}
    \item \textbf{0-shot}: a model that predicts emotion text given an instruction text and speech features defined as Eq. (15).
    \item $\textbf{1-shot}_N$: a model that predicts emotion text for target speech by conditioning a single neutral speech of a target speaker and a text described as ``Neutral''. 
\end{itemize}

\subsection{Setup}
These models were composed under the following conditions:
For the speech encoder, we used 80 log mel-scale filterbank coefficients as speech features. The frame shift was 10 ms. The speech features passed two convolution and max pooling layers with a slide of 2, so we down-sampled them to $1/4$ along with the time axis.
After these layers, we stacked 6-layer transformer encoder blocks.
For the Q-Former, we stacked 2-layer transformer decoder blocks without a causal attention mask.
Based on preliminary experiments, the length of the query was set to 150, as it yielded the best performance.
For the LLM encoder and decoder, we used 1024-dimensional token embeddings where the vocabulary size was set to 37,156 and stacked 32-layer transformer encoder blocks and 6-layer transformer decoder blocks. 
In the speech encoder, Q-Former, and LLM encoder and decoder, the dimensions of the output continuous representations were set to 512, 512, and 1024, those of the inner outputs in the position-wise feed-forward networks were set to 2048, 2048 and 4096, and the number of heads in the multi-head attention was set to 8, 8, and 16, respectively.
For training, we used the RAdam optimizer. We set the mini-batch size to 64 utterances and the dropout rate in the transformer blocks to 0.1. We introduced label smoothing, where its smoothing parameter was set to 0.1.
For inference with speech LM, we used a beam search algorithm in which the beam size was set to 4, and conditioned 0--7-shot enrollment utterance-label pairs for ICL.

\subsection{Evaluation}
As evaluation metrics, we used unweighted accuracy for the speaker ($\text{UA}_{\rm spk}$), which is the average of individual speaker accuracies, to evaluate whether personalized SER is performed for unseen speakers.
In addition, we calculated standard statistics, such as standard deviation, median, maximum, and minimum of $\text{UA}_{\rm spk}$.
Note that when evaluating the speech LM, which is a generative model, a generated result is correct if it exactly matches the correct answer.

\begin{table*}[tb]
  \centering
  \caption{Results of SER performance using our SER dataset.}
  \begin{tabular}{lccrrrrr}
      \hline
       Model & ICL & Setting & $\text{UA}_{\rm spk}$ ($\uparrow$) & $\sigma$($\text{UA}_{\rm spk}$) ($\downarrow$) & $\mu_{1/2}$($\text{UA}_{\rm spk}$) ($\uparrow$) & max($\text{UA}_{\rm spk}$) ($\uparrow$) & min($\text{UA}_{\rm spk}$) ($\uparrow$) \\
      \hline 
      Scratch & -- & -- & 0.482 &  0.099 & 0.473 & 0.711 & 0.277 \\
      Pretrained & -- & -- & 0.550 & 0.114 & 0.559 & 0.854 & 0.289 \\
      $\text{Personalized}_N$ \cite{9428217} & -- & -- & 0.595 & 0.108 & 0.611 & 0.846 & 0.346  \\
      $\text{Personalized}_A$ \cite{triantafyllopoulos2024} & -- & -- & 0.626 & 0.109 & 0.633 & 0.866 & 0.346 \\
     \hline 
     0-shot & -- & -- & 0.675 & 0.110 & 0.684 & 0.874 & 0.403 \\
     $\text{1-shot}_N$ & 1-shot & Neutral & 0.680 & 0.102 & 0.690 & 0.897 & 0.400 \\
     \hline
      Proposed & 0-shot & -- & 0.664 & 0.113 & 0.661 & 0.900 & 0.374\\
      & 1-shot & TU+LD & 0.698 & 0.099 & 0.707 & 0.911 & 0.411 \\
      & 2-shot & TU+LD & 0.717 & 0.102 & 0.723 & 0.926 & 0.420 \\
      & 3-shot & TU+LD & 0.732 & 0.097 & 0.739 & 0.949 & 0.446 \\
      & 4-shot & TU+LD & 0.738 & \bf{0.096} & 0.746 & 0.943 & 0.426 \\
      & 5-shot & TU+LD & 0.747 & 0.098 & 0.763 & 0.954 & 0.454 \\
      & 6-shot & TU+LD & 0.752 & 0.099 & 0.760 & \bf{0.960} & 0.460 \\
      & 7-shot & TO+LD & \bf{0.757} & 0.098 & \bf{0.766} & 0.957 & \bf{0.471} \\
      \hline
  \end{tabular}
  \label{table:main}
\end{table*}

\begin{table*}[tb]
  \centering
  \caption{Results of ICL-based inference using various ICL settings.}
  \begin{tabular}{lrrrrrrrrrrr}
      \hline
       ICL & TU+LD & TU+LU & TE+LD & TE+LU & \multicolumn{7}{c}{TU+LO} \\
       & & & & & Anger & Disgust & Fear & Joy & Sad & Surprise & Neutral \\
      \hline 
      1-shot & 0.698 & 0.698 & 0.692 & 0.692 & 0.695 & 0.696 & 0.704 & 0.699 & 0.702 & 0.703 & 0.690 \\
      2-shot & 0.717 & 0.715 & 0.706 & 0.708 & 0.705 & 0.703 & 0.712 & 0.711 & 0.709 & 0.711 & 0.696 \\
      3-shot & 0.732 & 0.726 & 0.718 & 0.714 & 0.709 & 0.705 & 0.717 & 0.716 & 0.711 & 0.716 & 0.700 \\
      4-shot & 0.738 & 0.734 & 0.726 & 0.721 & 0.712 & 0.708 & 0.716 & 0.721 & 0.710 & 0.720 & 0.699 \\
      5-shot & 0.747 & 0.740 & 0.734 & 0.727 & -- & -- & -- & -- & -- & -- & -- \\
      6-shot & 0.752 & 0.744 & 0.740 & 0.727 & -- & -- & -- & -- & -- & --  \\
      \hline
  \end{tabular}
  \label{table:various}
\end{table*}

\begin{table*}[tb]
  \centering
  \caption{Results of ICL-based inference using TU+LD and TU+LO settings.}
  \begin{tabular}{llrrrrrrr}
      \hline
       Test setting & Training setting & 1-shot & 2-shot & 3-shot & 4-shot & 5-shot & 6-shot & 7-shot   \\
       \hline 
       TU+LD & $\text{TU+LU}_{0:7}$  & \bf{0.698} & \bf{0.717} & \bf{0.732} & \bf{0.738} & \bf{0.747} & \bf{0.752} & 0.757 \\
       & $\text{TO+LD}_{7}$  & 0.563 & 0.621 & 0.659 & 0.659 & 0.721 & 0.746 & \bf{0.764} \\
       & $\text{Neutral}_{0:7}$  & 0.639 & 0.638 & 0.640 & 0.643 & 0.643 & 0.642 & 0.642 \\
      \hline
      TU+LO & $\text{TU+LU}_{0:7}$ & \bf{0.690} & \bf{0.696} & \bf{0.700} & \bf{0.699} & -- & -- & -- \\
      (Neutral) & $\text{TO+LD}_{7}$ & 0.516 & 0.512 & 0.508 & 0.506 & -- & -- & -- \\
      & $\text{Neutral}_{0:7}$ & 0.677 & 0.684 & 0.688 & 0.689 & -- & -- & -- \\
      \hline
  \end{tabular}
  \label{table:meta-training}
\end{table*}

\subsection{Results}
\paragraph*{Main result}
Table \ref{table:main} shows $\text{UA}_{\rm spk}$, standard deviation ($\sigma(\cdot)$), median ($\mu_{1/2}(\cdot)$), maximum (max($\cdot$)), and minimum (min($\cdot$)) values of $\text{UA}_{\rm spk}$ in each model.
The ``Neutral'' in the setting column of the table shows the performance when neutral utterances were only used in ICL-based inference. 
In the table, although the $\text{Personalized}_A$, which uses enrollment speech of all emotions, outperformed other models in the classifier model, it performed worse than the 0-shot.
This result indicates that the speech LM can learn the SER task better than classifier models.
In addition, in the proposed method, the SER performance improved as the number of enrollment utterance-label pairs increased, and the highest performance was achieved when all emotional utterances were used.
These results indicate that the proposed method can perform personalized SER via ICL.

\paragraph*{How to select enrollment utterances in ICL}
To investigate the impact of selecting methods of enrollment utterances on performance, we performed ICL using enrollment utterances that satisfy the following settings: TU+LD, TU+LU, TE+LD, TE+LU, and TU+LO settings.
Table \ref{table:various} shows the results $\text{UA}_{\rm spk}$ of each ICL setting.
In the table, using 1--2-shot examples achieved equivalent performance under all settings.
On the other hand, in the TU+LO setting, the ICL performance was improved or decreased slightly when using three or more examples. 
These results indicate that if the same emotion is used in ICL, up to two shots are effective for personalization.
In other settings, SER performance was improved with increased examples in ICL, and performance was improved in the order of TU+LD$>$TU+LU$>$TE+LD$>$TE+LU.
These results show that in SER via ICL, personalization works well when utterances can be prepared under TU or LU settings.

\paragraph*{How to select enrollment utterances in MetaICL}
To investigate the effects of the types of enrollment utterances used in MetaICL, we prepared two meta-trained speech LM using enrollment utterances that satisfy the following settings: $\text{TO+LD}_{7}$, which had only 7-shot examples using each emotion, and $\text{Neutral}_{0:7}$, which had 0--7-shot examples using only neutral utterance.
Table \ref{table:meta-training} shows the results of ICL-based inference using TU+LD and TU+LO (neutral utterances only) settings.
In the TU+LD setting, the performance of the speech LM meta-trained with $\text{TO+LD}_{7}$ was worse than that with $\text{Neutral}_{0:7}$ when 1--6-shot examples were used.
In the TU+LO setting, the performance of the speech LM meta-trained with $\text{TO+LD}_{7}$ decreased with increased examples.
These results show that the speech LM meta-trained with $\text{TO+LD}_{7}$ was optimized for input of 7-shot examples, so it cannot personalize SER via ICL.
Also, in the TU+LO setting,  the performance of the speech LM meta-trained with $\text{Neutral}_{0:7}$ was worse than that with $\text{TU+LU}_{0:7}$.
From the above, it is inferred that the speech LM can improve the personalization capability via ICL by meta-training using enrollment utterances expressing various emotions.

\section{Conclusion}
This paper proposed a method for personalizing speech emotion recognition (SER) via in-context learning (ICL).
In the proposed method, the speech language model (speech LM), extended from a large language model with ICL capability, performs personalization for SER by conditioning a few enrollment utterance-label pairs uttered by a target speaker in ICL-based inference.
Experimental results using our created SER dataset demonstrated that the proposed method had higher performance than conventional personalization methods, and SER performance was improved when utterances with arbitrary emotion labels were increased in ICL-based inference.

\end{document}